\documentclass[usenatbib]{mn2e}

\input epsf

\usepackage{graphicx}
\usepackage{txfonts}
\usepackage{epsfig}

\title[solarFLAG hare and hounds]{solarFLAG hare and hounds:
estimation of p-mode frequencies from Sun-as-star helioseismology
data}

\author[Jim\'enez-Reyes et
al.]{S. J. Jim\'enez-Reyes$^{1,2}$\thanks{E-mail: sjimenez@ll.iac.es},
W.~J.~Chaplin$^{2}$\thanks{E-mail: w.j.chaplin@bham.ac.uk},
R.~A.~Garc\'\i a$^3$, T.~Appourchaux$^4$,\newauthor F.~Baudin$^4$,
P.~Boumier$^4$, Y.~Elsworth$^2$, S.~T.~Fletcher$^5$,
M.~Lazrek$^{6,7,8}$,\newauthor J.~W.~Leibacher$^{9,4}$,
J.~Lochard$^1$, R.~New$^5$, C.~R\'egulo$^{1,10}$,
D.~Salabert$^{1,9}$,\newauthor T.~Toutain$^{2}$, G.~A.~Verner$^{11,2}$
and R.~Wachter$^{12}$\\ $^1$ Instituto de Astrof\'\i sica de Canarias,
38205, La Laguna, Tenerife, Spain\\ $^2$School of Physics and
Astronomy, University of Birmingham, Edgbaston, Birmingham, B15 2TT,
UK\\ $^3$ Laboratoire AIM, CEA/DSM-CNRS-Universit\'e Paris Diderot;
CEA-Saclay, IRFU, SAp, F-91191, Gif-sur-Yvette, France\\ $^4$Institut
d'Astrophysique Spatiale (IAS), B\^atiment 121, F-91405, Orsay Cedex,
France\\ $^5$Faculty of Arts, Computing, Engineering and Sciences,
Sheffield Hallam University, Sheffield S1 1WB, UK\\ $^6$LPHEA,
Facult\'e des Sciences Semlalia, Universite Cadi Ayyad, Marrakech,
Morocco\\ $^7$ D\'epartement d'Astrophysique, UMR 6525, Universit\'e
de Nice-Sophia Antipolis, 06108 Nice Cedex 2, France\\ $^{8}$
D\'epartement Cassini, URA CNRS 1362, Observatoire de la C\^ote
d'Azur, F-06304 Nice, France\\ $^{9}$National Solar Observatory, 950
North Cherry Avenue, Tucson, AZ 85719, USA\\ $^{10}$Dpto. de
Astrof\'isica, Universidad de La Laguna, La Laguna, 38206, Tenerife,
Spain\\ $^{11}$ Astronomy Unit, Queen Mary, University of London, Mile
End Road, London, E1 4NS, UK\\ $^{12}$ W. W. Hansen Experimental
Physics Laboratory, Stanford University, Stanford, CA 94305-4085\\}

\begin{document}

\maketitle

\begin{abstract}

We report on the results of the latest solarFLAG hare-and-hounds
exercise, which was concerned with testing methods for extraction of
frequencies of low-degree solar p modes from data collected by
Sun-as-a-star observations.  We have used the new solarFLAG simulator,
which includes the effects of correlated mode excitation and
correlations with background noise, to make artificial timeseries data
that mimic Doppler velocity observations of the Sun as a star. The
correlations give rise to asymmetry of mode peaks in the frequency
power spectrum. Ten members of the group (the hounds) applied their
``peak bagging'' codes to a 3456-day dataset, and the estimated mode
frequencies were returned to the hare (who was WJC) for
comparison. Analysis of the results reveals a systematic bias in the
estimated frequencies of modes above $\approx 1.8\,\rm mHz$. The bias
is negative, meaning the estimated frequencies systematically
underestimate the input frequencies.

We identify two sources that are the dominant contributions to the
frequency bias. Both sources involve failure to model accurately
subtle aspects of the observed power spectral density in the part
(window) of the frequency power spectrum that is being fitted.  One
source of bias arises from a failure to account for the power spectral
density coming from all those modes whose frequencies lie outside the
fitting windows. The other source arises from a failure to account for
the power spectral density of the weak $l=4$ and 5 modes, which are
often ignored in Sun-as-a-star analysis. The Sun-as-a-star
peak-bagging codes need to allow for both sources, otherwise the
frequencies are likely to be biased.

\end{abstract}


\section{Introduction}

The solar Fitting at Low-Angular degree Group
(\emph{solarFLAG})\footnote{See
http://bison.ph.bham.ac.uk/\,\~\,wjc/Research/FLAG.html} has as its
main aims the development and refinement of techniques for analysis of
data from the low-degree (low-$l$) p modes, from observations made of
the ``Sun as a star'' . These data play a crucial r\^ole in studies of
the deep radiative interior and core of the Sun.

The input data for probing the solar interior are the mode parameters,
such as individual frequencies, frequency splittings, damping rates
and powers. The mode frequencies may be used to infer the internal
hydrostatic structure (sound speed, density); accurate and precise
frequencies are a vital prerequisite for ensuring that robust
inference is made on the structure.

Analysis of the Sun-as-a-star (and also the resolved-Sun) helioseismic
data requires application of complicated algorithms to extract
estimates of the mode parameters. This usually involves fitting
multi-parameter models to the resonant peaks in the frequency power
spectrum of the observations. An important aim of the solarFLAG
program is to quantify levels of bias arising from, and precision
achievable in, these \emph{peak bagging} procedures. ``Hare and
hounds'' exercises on realistic artificial data form the framework for
this activity.

In a first study (Chaplin et al. 2006) we looked in detail at the
accuracy and precision of rotational frequency splittings extracted
from a 3456-d set of artificial Sun-as-a-star data, to which ten
members of the solarFLAG applied their peak-bagging codes. The
parameters we look at in this paper are the low-$l$ mode frequencies
returned by the peak-bagging codes.  The sets of artificial timeseries
data used in our first study did not include any asymmetry of the
simulated p-mode peaks in the frequency power spectrum; this asymmetry
is exhibited by the real solar p modes. The peak-bagging codes must be
able to cope with the asymmetry to in principle allow accurate
estimation of the mode frequencies. It was therefore clear to us that
to expedite a meaningful hare-and-hounds study on the mode frequencies
we would need to generate artificial data with asymmetry
included. This we have now done, and this paper reports on results of
a hare-and-hounds exercise conducted with the new asymmetric
artificial data, to which ten members of the solarFLAG applied their
peak-bagging codes.

We have used a simple, but very powerful method to introduce in the
time domain the effects of asymmetry, which is based on the framework
proposed by Toutain, Elsworth \& Chaplin (2006). There are two main
factors in the method that contribute to the asymmetry of the
artificial mode peaks. First, background noise is correlated with the
excitation functions of the modes. Second, overtones of the same
angular degree and azimuthal order have excitation functions that are
correlated in time (see Chaplin, Elsworth \& Toutain 2008). In this
framework, correlation of the excitation follows naturally from
invoking correlations with the background noise. As we shall see in
this paper, the power spectral density of the resulting asymmetric
mode peaks must be modelled accurately, otherwise estimates of the
mode frequencies returned by the peak-bagging codes will be biased.
This is the main result of the paper.

The layout of the rest of the paper is as
follows. Section~\ref{sec:simu} gives a brief overview of the
solarFLAG simulator, which was used to make the artificial timeseries
data for the hare-and-hounds exercise.  We also discuss in this
section the basic attributes of the artificial dataset analyzed by the
ten hounds. A detailed description of the simulator, which pays
particular attention to the impact of correlations in the data, is
given in Chaplin et al. (in preparation).

Section~\ref{sec:fit} summarizes the main elements of the fitting
strategies that were adopted by the hounds. Section~\ref{sec:res} then
presents the main results of the hare-and-hounds exercise. We look in
detail at how the estimated frequencies of the hounds compared, not
only against the input frequencies (results which bear on the accuracy
of the peak-bagging procedures), but also against one another (results
which bear on the precision inherent in the estimated frequencies). In
Section~\ref{sec:disc} we identify the origins of a systematic
frequency bias that is reported in Section~\ref{sec:res}.  Finally, we
conclude in Section~\ref{sec:sum} with a summary of the main
conclusions of the paper, where we also discuss implications of the
frequency bias for the fitting strategies.

 \section{The solarFLAG simulator}
 \label{sec:simu}

 \subsection{General information}
 \label{sec:gensimu}

The solarFLAG datasets simulate full-disc Sun-as-a-star Doppler
velocity observations, such as those made by the ground-based
Birmingham Solar-Oscillations Network (BiSON) and the Global
Oscillations at Low-Frequency (GOLF) instrument on board the
\emph{ESA/NASA} SOHO spacecraft.  The dataset made by the hare (who
was WJC) for the hare-and-hounds exercise spanned 3456 simulated days,
with data samples made on a regular 40-sec cadence. The dataset did
not include any solar-cycle effects. The impact of these effects will
be dealt with in a separate paper.

solarFLAG datasets are made with a full complement of simulated
low-$l$ modes. The hare-and-hounds dataset included all modes in the
ranges $0 \le l \le 5$ and $1000 \le \nu \le 5000\,\rm \mu
Hz$. Frequencies of the modes came from standard solar model BS05(OP)
of Bahcall et al. (2005). We also added a surface term to these
frequencies, which was based on polynomial fits to differences between
the raw model BS05(OP) frequencies and frequencies from analysis of
BiSON and GOLF data. A database of p-mode power, linewidth and peak
asymmetry estimates, obtained from analyses of GOLF and BiSON data,
was used to guide the choice of the other input mode parameters.

The hypothetical solarFLAG instrument was assumed to make its
observations from a location in, or close to, the ecliptic plane.
This is the perspective from which BiSON (ground-based network) and
GOLF (orbiting the Sun at the L1 Lagrangian point) view the Sun. The
rotation axis of the Sun is then always nearly perpendicular to the
line-of-sight direction, and only a subset of the $2l+1$ components of
the non-radial modes are clearly visible: those having even $l+m$,
where $m$ is the azimuthal order. These components are represented
explicitly in the solarFLAG timeseries. The visibility for given
($l$,\,$m$) also depends, though to a lesser extent, on the spatial
filter of the instrument (e.g., see Christensen-Dalsgaard 1989).
Here, we adopted BiSON-like visibility ratios.

We included two sources of background noise in the data, which have a
significant power contribution in the range occupied by the p
modes. First, a simple photon shot noise component, having a white
frequency power spectrum. This component was made in the time domain
from random Gaussian noise, specified by a sample standard deviation
of $\sigma_{\rm psn} = 0.25\,\rm m\,s^{-1}$ per 40-sec sample. The
other source of background was granulation-like noise, having a
frequency power spectrum like the Harvey (1985) power-law model. As we
shall see below -- where we also specify its basic parameters -- this
noise is used to excite the modes, and plays a crucial r\^ole in the
correlations introduced in the data.

 \subsection{Correlated excitation and correlated noise}
 \label{sec:corr}

The solarFLAG simulator models the effects of correlated mode
excitation and correlated background noise. One does not need to
understand the detail of the implementation in order to follow the
discussion of the results in this paper. Rather, one needs only to
take away the two following, key points: First, inclusion of
correlation effects gives rise to asymmetric peaks in the frequency
power spectrum; and these correlations may be tuned in the simulator
to give asymmetries which resemble closely those displayed in real
Sun-as-a-star data.  Second, the impact of the correlations is such
that the power spectral density in the outlying tails of the mode
peaks falls off in a manner that is evidently similar to that in the
real data. Frequency power spectra of full solarFLAG timeseries
therefore show a close resemblance, in their overall appearance, to
real Sun-as-a-star spectra. We considered these two points as
important baseline requirements for any artificial dataset used in the
hare-and-hounds exercises.

An in-depth discussion of the implementation of the correlations, and
full details on the simulator, are given in Chaplin et al. (in
preparation). In what remains of this section we give a summary of the
basic principles, and a brief overview of the simulator. We also show
the underlying peak asymmetries that were introduced in the solarFLAG
hare-and-hounds dataset.

 \subsubsection{Basic principles}
 \label{sec:basic}

In Toutain, Elsworth \& Chaplin (2006) it was hypothesized that the
excitation function of an overtone, $n$, with angular degree $l$ and
azimuthal degree $m$ (whose frequency is $\nu_{nlm}$), is the same as
that component of the solar background (granulation) noise that has
the same spherical harmonic projection, $Y_{lm}$, in the corresponding
range in frequency in the Fourier domain. An important implication is
that overtones with the same ($l$,\,$m$) should have excitation
functions that are correlated in time. (Note that the $Y_{lm}$ for
($l$,\,$m$) and ($l$,\,$-m$) are orthogonal, and are therefore assumed
to have independent, i.e., uncorrelated, excitation.) Moreover, since
Doppler velocity observations of the Sun are also sensitive to the
granulation background, perturbations due to the modes and this noise
will be correlated in time. This is what we call correlated background
noise (see also, e.g., Roxburgh \& Vorontsov 1997; Severino et
al. 2001; Gabriel et al. 2001; Jefferies et al. 2003; Barban, Hill \&
Kras 2004; and references therein).

Even in the absence of any correlated background noise, the correlated
excitation would give rise to asymmetry of mode peaks in the frequency
power spectrum. This asymmetry is due to complex interactions between
the tails of the correlated mode peaks. When the correlated background
noise is included, there are then additional contributions to the peak
asymmetry.

 \subsubsection{Inclusion of correlation effects in the simulator}
 \label{sec:incl}

The basis of the solarFLAG simulator is the method discussed in
Chaplin et al. (1997) for generating timeseries of individual p
modes. The method uses the Laplace transform solution of the equation
of a forced, damped harmonic oscillator to make the output velocity of
each artificial mode. Oscillators are re-excited at each time sample
-- the chosen cadence here being 40\,sec -- with small `kicks' from a
timeseries of random noise.

The kicks are drawn from a timeseries of granulation-like noise. This
noise is made by using random white-noise input to a low-order,
autoregressive model. Overtones of a given ($l$,\,$m$) have kicks that
are correlated in time. The granulation-like noise is also used to
give the correlated background noise. The granulation-like noise is
specified by two free parameters: $\sigma$ fixes the amplitude; while
the characteristic timescale, $\tau$, is given a value to mimic the
lifetime of granules on the Sun.

A single constant, $\rho$, fixes the coefficient of correlation for
the correlated excitation, and the correlation with the background
(see also Chaplin, Elsworth \& Toutain 2008). This gives the user the
flexibility to ``tune'' the asymmetry of the mode peaks -- the higher
is $\rho$, the larger is the asymmetry. When $\rho=\pm 1$, overtones
with the same ($l$,\,$m$) are all excited by the same timeseries; when
$\rho=0$, they are excited by statistically independent timeseries;
and when $0< |\rho| < 1$ they are kicked by a mixture of common and
independent timeseries. The common timeseries (or a mixture of common
and independent timeseries data) is later added as background noise
(having been suitably scaled in amplitude first).

\begin{figure}
   \centerline{\epsfxsize=8.0cm\epsfbox{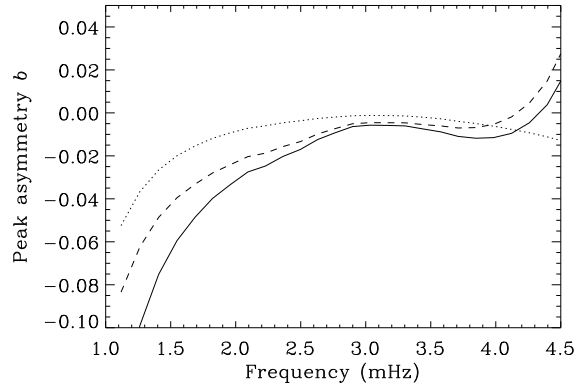}}

 \caption{Peak asymmetry in the solarFLAG hare-and-hounds
 dataset. Thick solid line: total asymmetry. Dotted line: contribution
 due to correlated background noise. Dashed line: contribution due to
 correlated mode excitation.}

 \label{fig:specass}
 \end{figure}

Fig.~\ref{fig:specass} shows the input asymmetry given to mode peaks
in the hare-and-hounds dataset.  We note that the solarFLAG simulator
was configured on the assumption that at a given frequency the
relative sizes of the granulation noise and the mode amplitudes are
independent of degrees $l$ and $m$. An important consequence of this
assumption is that, at a given frequency, the asymmetry contribution
from the correlated noise -- shown here as the dotted line -- is the
same for all ($l$,\,$m$). This contribution is fixed for a given mode
by three free parameters, the p-mode parameters (i.e., frequencies,
heights, linewidths) having already been fully specified on input. The
free parameters are: $\rho$, $\sigma$ and $\tau$. The hare-and-hounds
timeseries was made with $\sigma=0.2\,\rm m\,s^{-1}$ and
$\tau=260\,\rm sec$. The hare also settled on $\rho=-0.36$. Use of
negative $\rho$ gave negative peak asymmetry, as displayed in real
Sun-as-a-star data; while the absolute value of $\rho$ gave asymmetry
that matched reasonably well that seen in BiSON and GOLF data.


The dashed line in Fig.~\ref{fig:specass} shows the contribution to
the peak asymmetry arising from the correlated excitation. This
contribution is fixed by the frequency separations, linewidths and
relative heights of the overtones, and the choice of $\rho$. Since
these mode parameters are similar for all ($l$,\,$m$), so too are the
peak asymmetries from this contribution (the figure shows the
contribution for overtones of $l=0$). Again, full details on all of
the above are given in Chaplin et al. (in preparation).

The solid line in Fig.~\ref{fig:specass} shows the total input
asymmetry for the hare-and-hounds solarFLAG dataset, given by the
combined effect of the correlated noise and correlated
excitation\footnote{There is actually a third contribution to the peak
asymmetry, from the non-white frequency response of the
excitation. The response in the vicinity of each resonance of course
rises with decreasing frequency, meaning there will be a small
negative asymmetry contribution. This contribution is, however, very
small compared to the contributions from correlated noise and
correlated excitation.}. This is the asymmetry actually displayed by
peaks in the frequency power spectrum, and is the asymmetry the hounds
would aim to recover when fitting the asymmetry as a free parameter.

 \section{Fitting strategies of the hounds}
 \label{sec:fit}

Ten members of the solarFLAG acted as hounds. They applied their
peak-bagging codes to the frequency power spectrum of the complete
hare-and-hounds dataset to recover estimates of the artificial mode
frequencies. The ten hounds were: PB, STF, RAG, SJJ-R, ML, JL, DS, TT,
GAV and RW. A priori information given to the hounds was limited to:
the cadence and length of the timeseries; and the calibration and
format of the stored residuals. For the purposes of this study we
chose not to impose an observational window function on the timeseries
(e.g., that from a ground-based network). This meant parameter
extraction was tested under the more favourable conditions afforded by
a 100-per-cent duty cycle.

All hounds adopted a peak-bagging approach to the analysis. We refer
the reader back to Chaplin et al. (2006) for more
details. Peak-bagging involves maximum-likelihood fitting of mode
peaks in the frequency power spectrum to multi-parameter fitting
models, where individual mode peaks are represented by Lorentzian-like
functions. Here, all hounds used the asymmetric Lorentzian-like
formula of Nigam \& Kosovichev (1998) to model individual peaks.

A common peak-bagging strategy is to go through the frequency power
spectrum fitting a mode pair at a time (the so-called ``pair-by-pair''
approach). This is because the $l=0$ modes lie in close proximity in
frequency to the $l=2$ modes. The same is true for the $l=1$ and $l=3$
modes. Eight hounds used this standard approach, isolating narrow
frequency windows, centred on the target pairs, to perform the
fitting.  Chosen window sizes varied from 40 to $50\,\rm \mu Hz$ for
the even-$l$ pairs, and 40 to $60\,\rm \mu Hz$ for the odd-$l$ pairs.

In the standard approach the fitting models usually only include power
from the target pair. They also use a flat offset to represent the
pseudo-white background (which varies only very slowly with frequency
in the range of interest). However, the models then fail to account
for power spectral density in the fitting window that comes from two
other sources: (i) the nearby, weak $l=4$ and 5 peaks; and (ii), the
slowly-decaying tails of the other even and odd-$l$ pairs in the
spectrum, whose resonant frequencies lie outside the fitting window.

The first of these sources is a bigger cause for concern where results
on the even-$l$ pairs are concerned, since the $l=4$ and 5 modes
usually lie in their fitting windows (e.g., see Fig.~8 of Chaplin et
al., 2006). This is not usually the case for the odd-$l$ pairs.  A few
hounds submitted results that allowed for the presence of the $l=4$
and 5 modes.

A way around problems caused by the second source is of course to take
account of the outlying power (e.g., see Jimen\'ez, Roca-Cort\'es \&
Jimen\'ez-Reyes 2002; Gelly et al. 2002; Fletcher et al. 2008), or to
fit all modes in the frequency power spectrum in one go (e.g., see
Lazrek et al. 2000; Appourchaux 2003). Two hounds also submitted
results where they allowed for the outlying power in their fitting
models. However, they did so by modelling the tails of the outlying
peaks as symmetric Lorentzians, not the asymmetric functions actually
displayed in the frequency power spectrum.

As we shall see in Section~\ref{sec:disc}, failure to deal properly
with sources (i) and (ii) leads to bias in the estimated frequencies,
and this dominates the other potential sources of bias. An example of
another candidate source was use of inaccurate mode-component
visibilities when modelling $l=2$ and 3 multiplets during
peak-bagging. In our first solarFLAG study (Chaplin et al., 2006), we
showed that this was a major source of bias for estimation of the
rotational frequency splittings. The frequencies are in contrast
remarkably insensitive to the choice of the visibilities. There were
differences in the numbers of parameters the hounds sought to estimate
by fitting, and we also checked whether these differences could have
affected the frequencies. One example was a division between those
hounds who constrained the widths of all components in both modes of a
pair to be the same (again, common practice at low $l$); and those who
instead fitted two widths, one for each mode. This strategy did not
have a significant impact on the estimated frequencies.

 \section{Results}
 \label{sec:res}


 \begin{figure*}
   \centerline{\epsfxsize=12.0cm\epsfbox{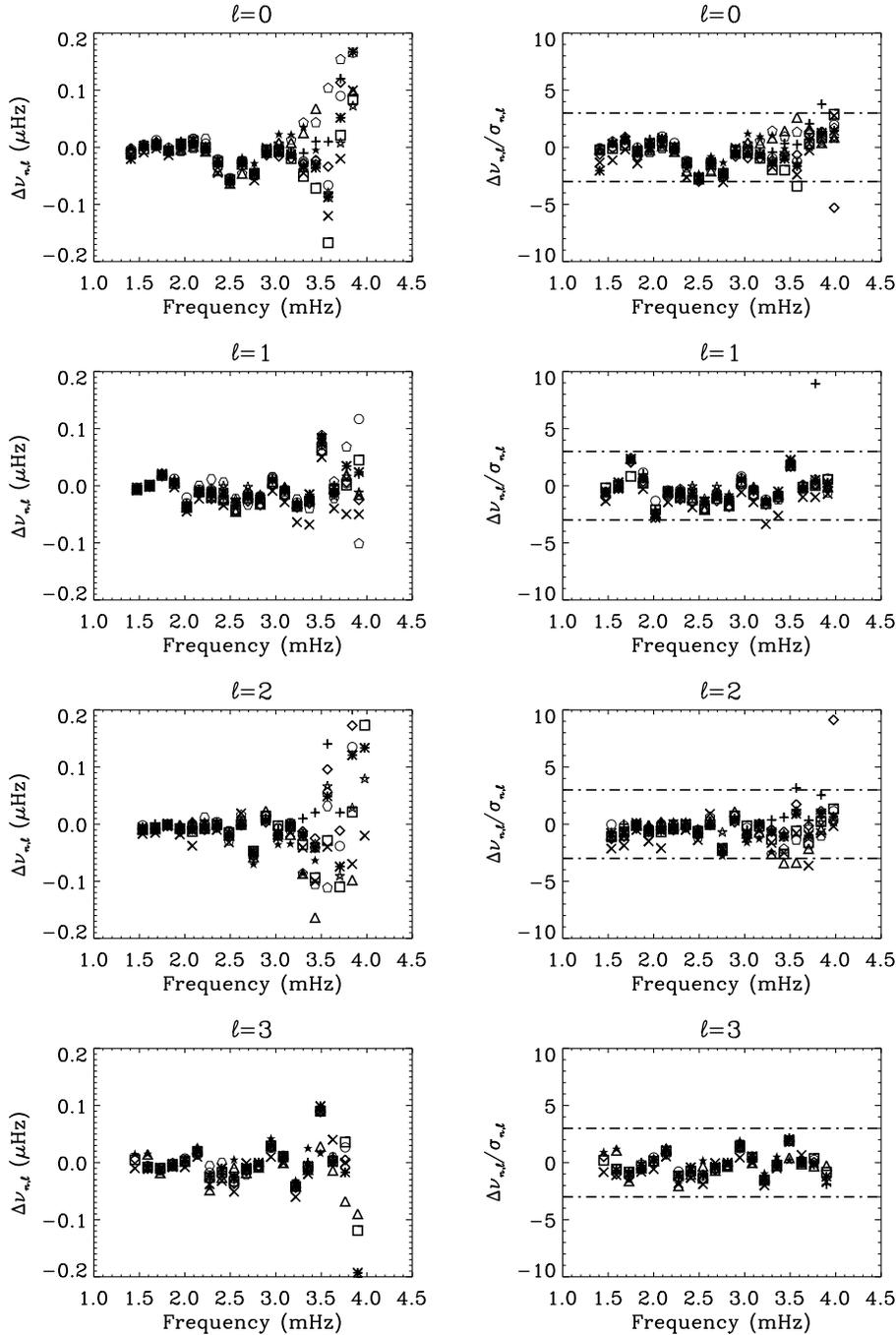}}

 \caption{Left-hand panels: differences between the fitted and input
 frequencies (in the sense fitted minus input) at each degree, $l$
 (different symbol for each hound). Right-hand panels: Differences in
 the left-hand panels normalized by the estimated frequency uncertainties,
 to give differences in units of sigma. The dot-dashed lines mark the
 $\pm 3\sigma$ levels.}

 \label{fig:res}
 \end{figure*}


 \begin{figure*}
	\centerline{\epsfxsize=5.0cm\epsfbox{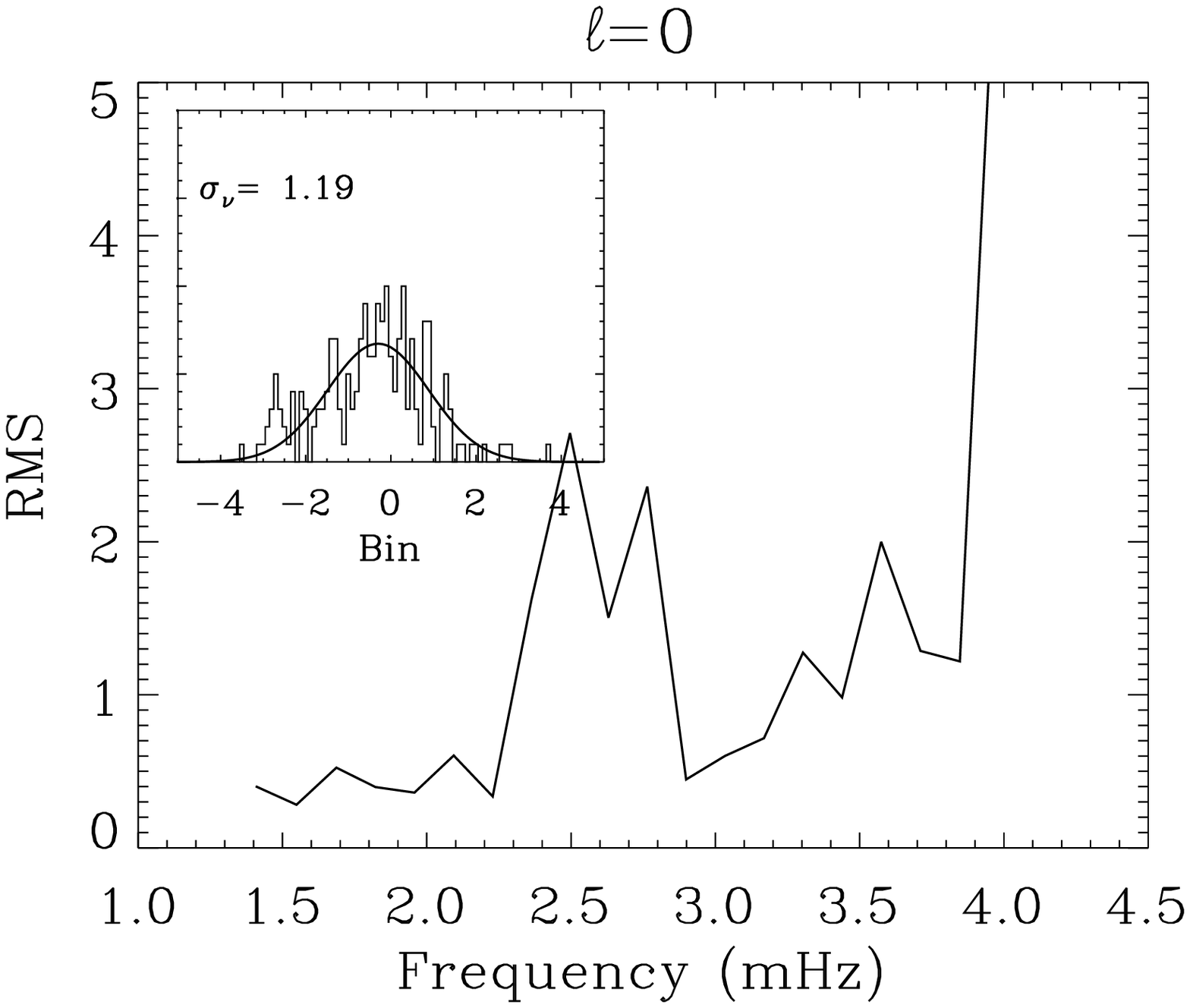}
        \hspace{1cm}\epsfxsize=5.0cm\epsfbox{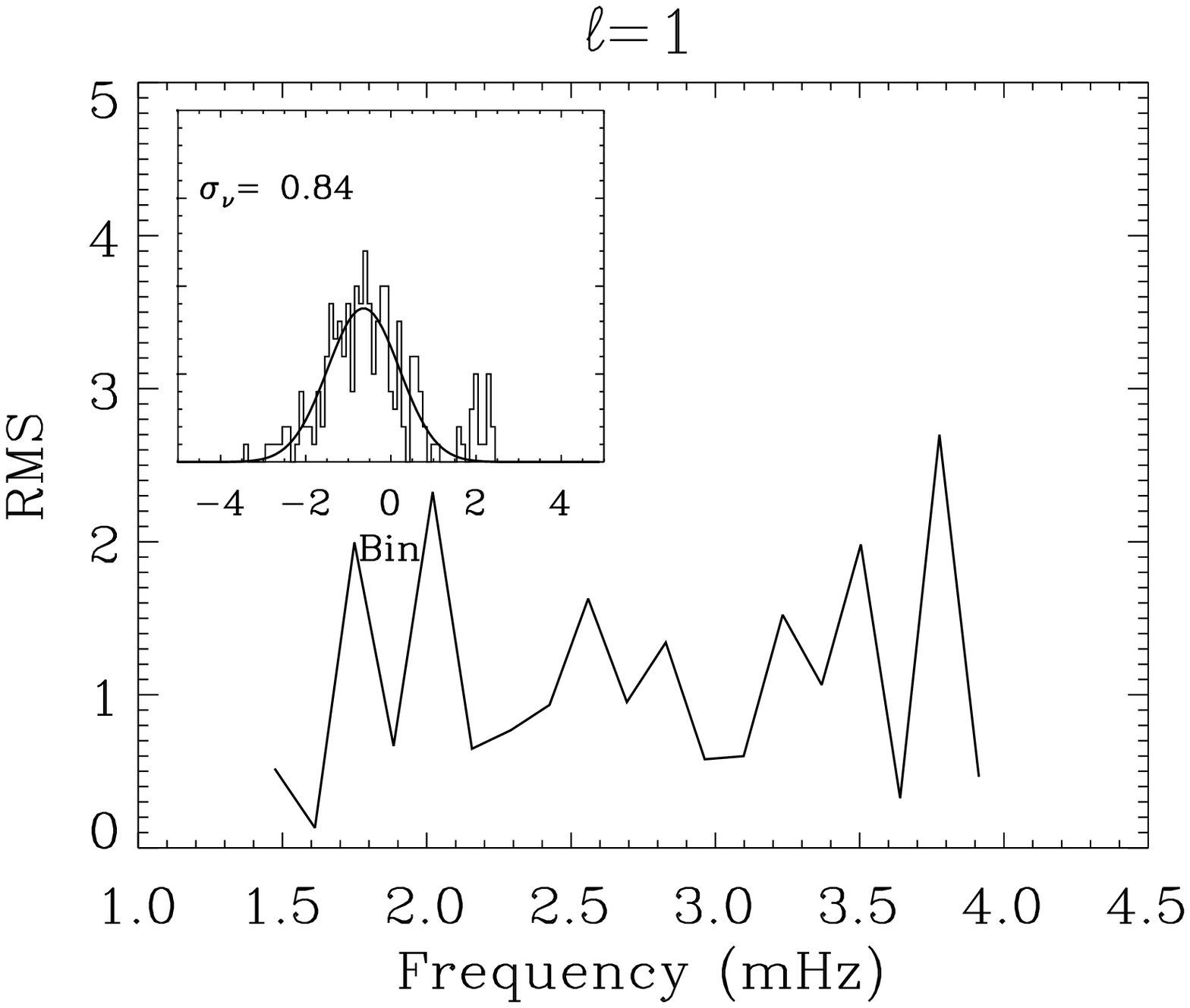}}
   	\vspace{.5cm}
   	\centerline{\epsfxsize=5.0cm\epsfbox{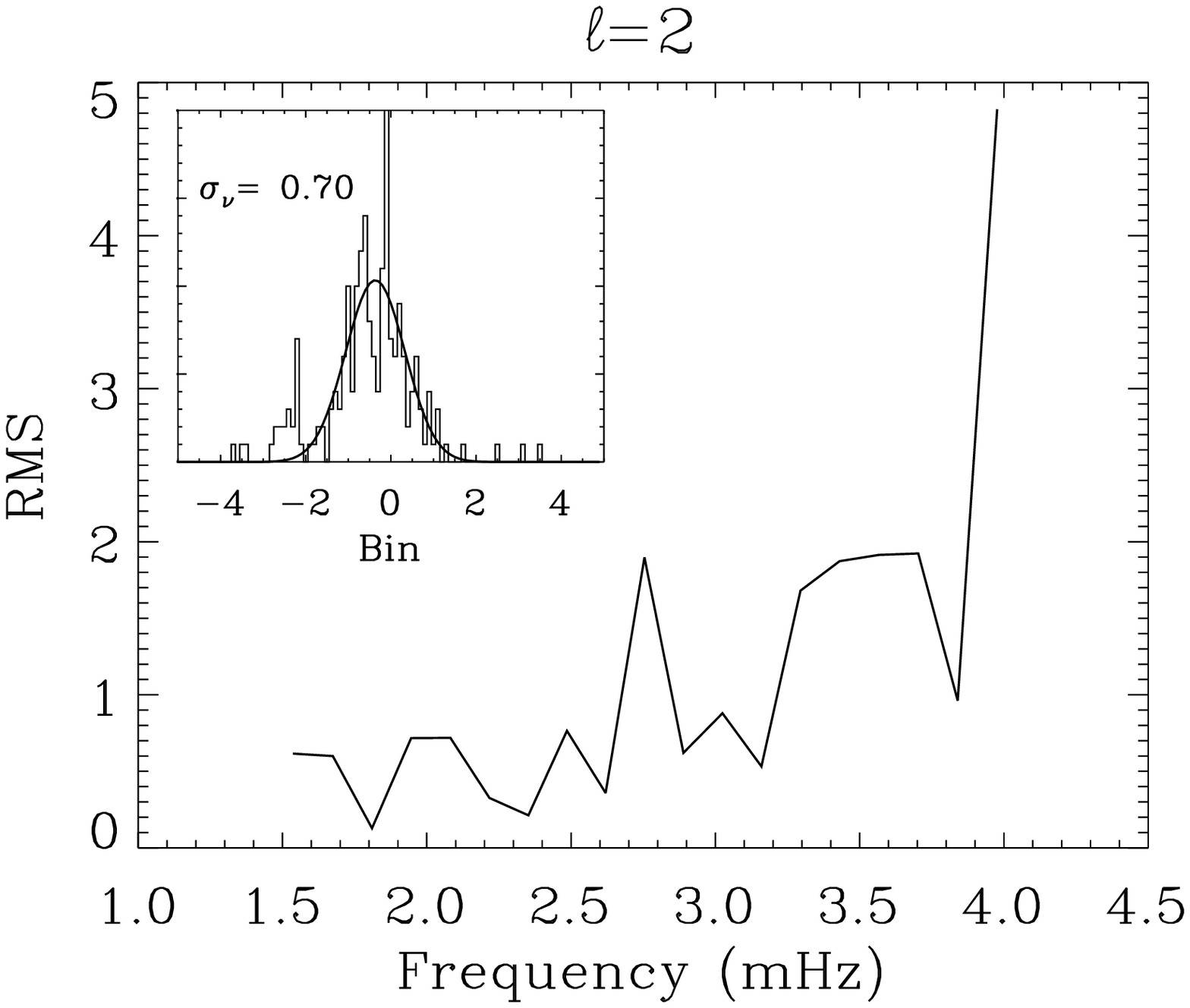}
        \hspace{1cm}\epsfxsize=5.0cm\epsfbox{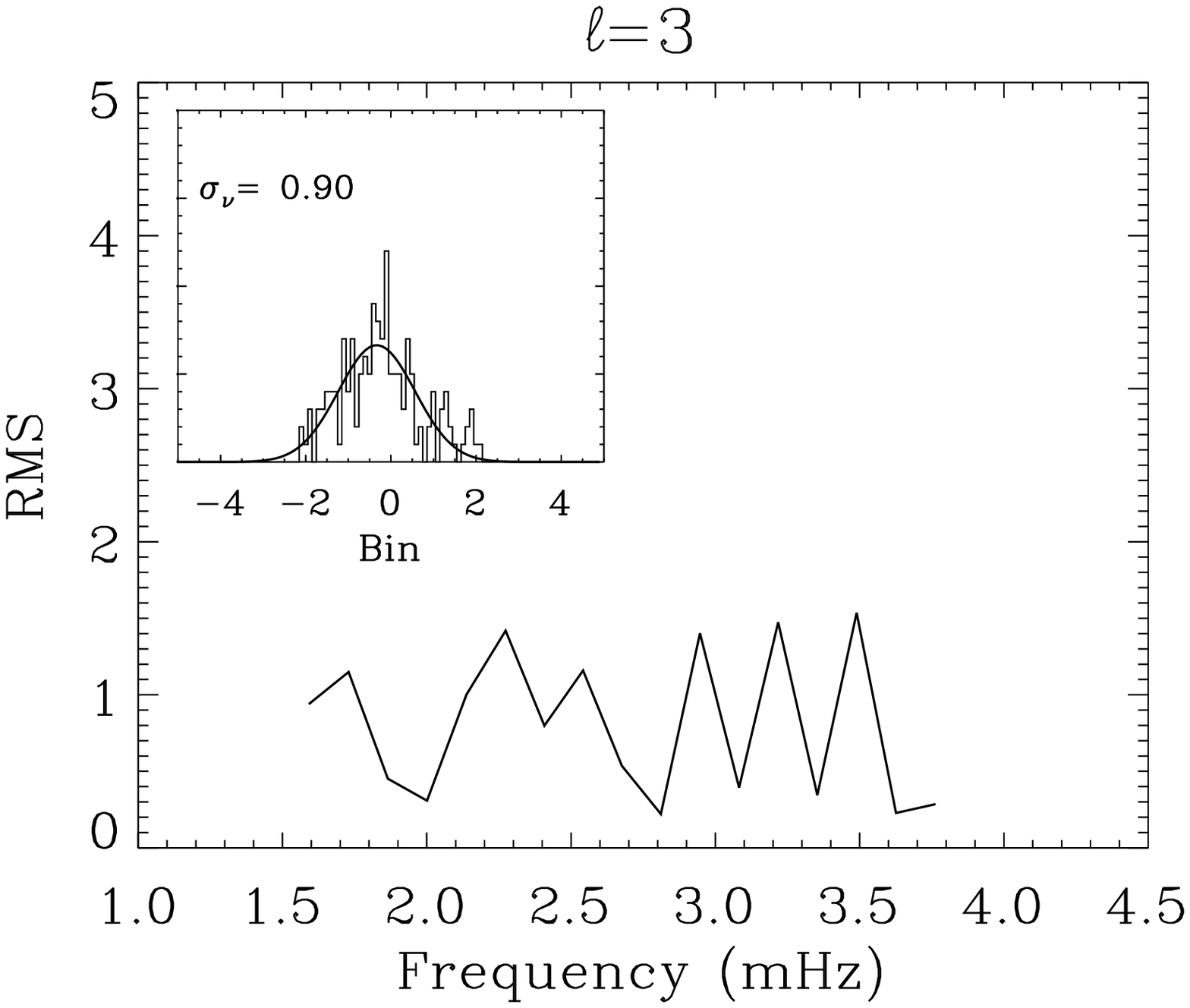}}

 \caption{The curves in each panel show the \textsc{rms} frequency
 differences, normalized by the mean of the fitting uncertainties
 returned by the hounds. These differences are referred to as
 normalized \textsc{rms} differences in the text. Insets show
 histograms of the normalized \textsc{rms} differences at each degree,
 $l$.}
  
 \label{fig:rms}
 \end{figure*}


The main results of the hare-and-hounds exercise are shown in
Figs.~\ref{fig:res} and~\ref{fig:rms}.

The results in Fig.~\ref{fig:res} bear on the accuracy of the fitted
frequencies. The four left-hand panels of this figure plot differences
between the fitted and input frequencies (in the sense fitted minus
input) at each degree, $l$. A different symbol is used to illustrate
the results of each hound. In order to give a direct measure of the
significance of these frequency differences, we divided the
differences by the estimated frequency uncertainties. All hounds
estimated uncertainties in the same way, taking, for each fit, the
square root of the appropriate diagonal element of the inverted
Hessian fitting matrix. The resulting normalized frequency differences
(units of sigma) are plotted in the four right-hand panels of
Fig.~\ref{fig:res}. The dot-dashed lines, which mark the $\pm
3\sigma$-levels, are included as eye guides.

What conclusions may we draw from the results in Fig.~\ref{fig:res}?
While at the lowest frequencies agreement between the fitted and input
frequencies is very good, over most of the fitting range there is a
persistent negative bias in the fitted frequencies. The significance
of this bias reaches $\approx 3\sigma$ for some of the modes, and is
largest at $l=0$ (e.g., see the $l=0$ results near $\approx 2.6\,\rm
mHz$). It is striking how the results of the different hounds follow
one another quite closely at frequencies below $\approx 3.1\,\rm
mHz$. This reflects the fact that all fits are affected by the same
realization noise. However, we shall show below in
Section~\ref{sec:disc} that the negative bias \emph{is not} simply a
consequence of the realization noise, and that fitting results on
timeseries made with the same input parameters, but different
realization noise, also show negative bias.

At high frequencies the fitted frequencies are more scattered. In this
part of the p-mode spectrum, large linewidths (high damping rates)
cause nearby peaks to overlap in frequency. This happens not only
within individual multiplets, where the effect becomes important above
$\approx 3\,\rm mHz$ in the closely spaced $l=1$ multiplets, but also
between adjacent modes in the low-$l$ pairs, where the effect becomes
important above $\approx 3.5\,\rm mHz$ for even-$l$ pairs (and at
higher frequencies for the more widely separated odd-$l$ pairs).

Next, let us consider differences \emph{between} the hounds.  These
differences bear on the precision of the results. That is because
disagreements in the results of fitting the same dataset imply the
frequencies are not as precisely (or, for that matter, accurately)
known as we might otherwise think. Disagreement in the results may be
thought of as an additional source of uncertainty for the estimated
frequencies, over and above that due to the stochastic excitation and
finite signal-to-noise ratio of the data.

This extra source of error -- often referred to as reduction noise --
may be estimated as follows.  For each mode we calculated an
\textsc{rms} frequency difference of the fitted frequencies of the
hounds, and then normalized that difference by the average of the
hounds' uncertainties for that mode, to give the normalized
\textsc{rms} differences plotted in Fig.~\ref{fig:rms} (units of
sigma).

When the normalized \textsc{rms} differences are close to zero, we may
infer that agreement between the hounds is excellent. However, the
results in Fig.~\ref{fig:rms} indicate this is clearly not the case
for many of the fitted modes, where the normalized \textsc{rms}
differences are comparable to, or even larger in size than, the
fitting uncertainties (i.e., the $1\sigma$ level). In order to get a
measure of the typical size of this extra uncertainty, we computed
histograms of the normalized \textsc{rms} differences at each degree,
$l$. These histograms are displayed as an inset to each panel of
Fig.~\ref{fig:rms}. The annotation also shows the standard deviations
of the best-fitting Gaussian profiles of each histogram, which are in
all cases comparable in size to $1\sigma$. We conclude that reduction
noise, arising from differences between the hounds, constitutes a
significant source of uncertainty for the estimated frequencies.


 \begin{figure*}
   \centerline{\epsfxsize=12.0cm\epsfbox{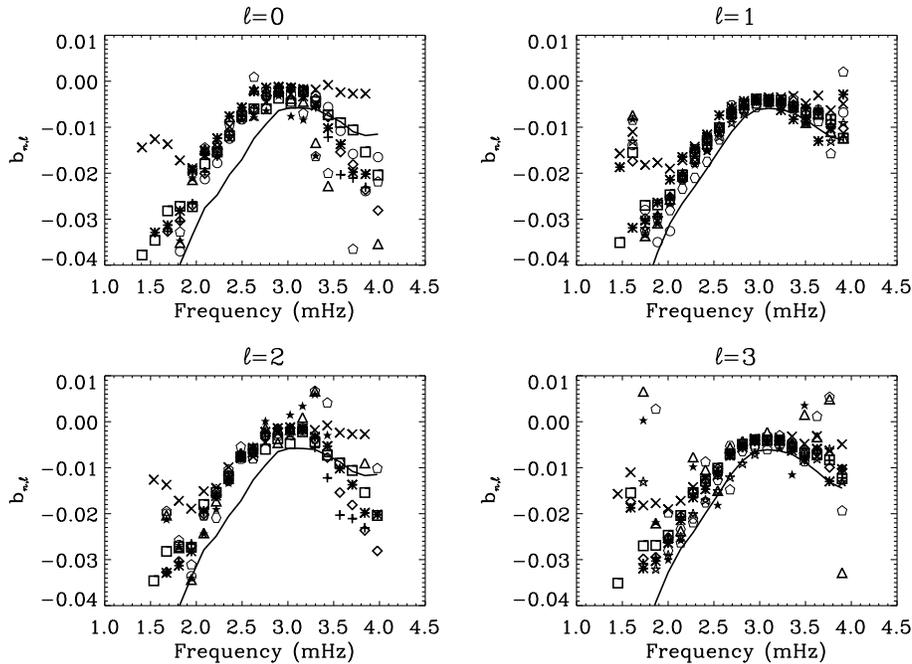}}

 \caption{The estimated peak asymmetries (different symbol for each
 hound). The solid line in each panel is the input asymmetry.}

 \label{fig:ass}
 \end{figure*}


Finally in this section, we also look at results on the fitted peak
asymmetries. Poor estimation of the asymmetries will bias the fitted
frequencies, and so results on the asymmetries are of considerable
interest. Fig.~\ref{fig:ass} shows the fitted asymmetries of the
hounds. The solid line in each panel is the input asymmetry (also
shown as the solid line in Fig.~\ref{fig:specass}). It is evident that
several poor estimates of the asymmetry were returned at the lowest
frequencies. Here, the height-to-background ratio of the peaks takes
its smallest values in the spectrum, and the peaks are also very
narrow, making determination of the asymmetry less straightforward
than in the main part of the spectrum. At the highest frequencies, the
overlap of peaks also presents difficulties for the analysis. However,
the most striking aspect of Fig.~\ref{fig:ass} is the persistent bias
present in the estimates over the main part of the spectrum, where the
returned estimates systematically underestimate the actual input size
of the asymmetry. We turn next to a detailed discussion of the bias.

 \section{Discussion of Results}
 \label{sec:disc}

It turns out that two sources give a significant contribution to the
bias in the frequencies shown in Section~\ref{sec:res}. Both sources,
which were noted previously in Section~\ref{sec:fit}, involve failure
to model accurately subtle aspects of the observed power spectral
density in the fitting windows. Again, they are: (i) power from $l=4$
and 5 modes, which affects fits to the even-$l$ pairs; and (ii) power
from the slowly-decaying tails of the other even and odd-$l$ pairs in
the spectrum, whose resonant frequencies lie outside the fitting
windows. In order to show clearly the bias from both of these sources,
and to thereby explain the results from Section~\ref{sec:res}, we
present here additional peak-bagging results. These results come from
fits made to many independent realizations of artificial solarFLAG
datasets. These datasets were identical, or similar, to the
hare-and-hounds dataset.

The hare made four sequences of data. Each sequence was comprised of
25 independent realizations of the same artificial Sun. The four
artificial suns defining each sequence had the same mode parameter,
granulation noise and shot noise characteristics as the
hare-and-hounds dataset. However, the coefficient describing the
correlation of the excitation and noise background, and the number of
degrees $l$ in the data, was varied from one sequence to another. The
content of the four sequences may be summarized as follows:\\

Sequence\,\#1 --- Datasets in this sequence were comprised of modes
from $l=0$ to 3, but there were \emph{no} $l=4$ and 5
modes. Furthermore, the excitation of the modes was uncorrelated; this
meant there was also no correlation with the granulation-like noise
background. The coefficient of correlation, $\rho$, was therefore set
to zero, and all peaks were symmetric Lorentzians.\\

Sequence\,\#2 --- Datasets in this sequence were comprised of modes
from $l=0$ all the way up to $l=5$. But, like Sequence\,\#1, there was
no correlation of the excitation, or correlation with the
granulation-like noise (so, again, $\rho=0$ and all peaks were again
symmetric).\\

Sequence\,\#3 --- Datasets in this sequence were comprised of modes
from $l=0$ to 3, with \emph{no} $l=4$ and 5 modes. However,
correlation of the excitation, and correlation with the
granulation-like noise background, was included. The coefficient of
correlation was given the same value as the hare-and-hounds dataset,
i.e., $\rho=-0.36$. The mode peaks were therefore asymmetric.\\

Sequence\,\#4 --- Datasets in this sequence had the same underlying
parameters as the hare-and-hounds dataset, i.e., modes up to $l=5$,
and correlations fixed by $\rho=-0.36$ (so the mode peaks were
asymmetric).\\

The hare then applied a standard (i.e., pair-by-pair) peak-bagging
code to the frequency power spectrum of each dataset. This standard
code fitted modes a pair at a time, and did not account for the $l=4$
and 5 modes, or outlying power from modes outside the fitting windows.

\begin{figure*}
   \centerline{\epsfxsize=7.0cm\epsfbox{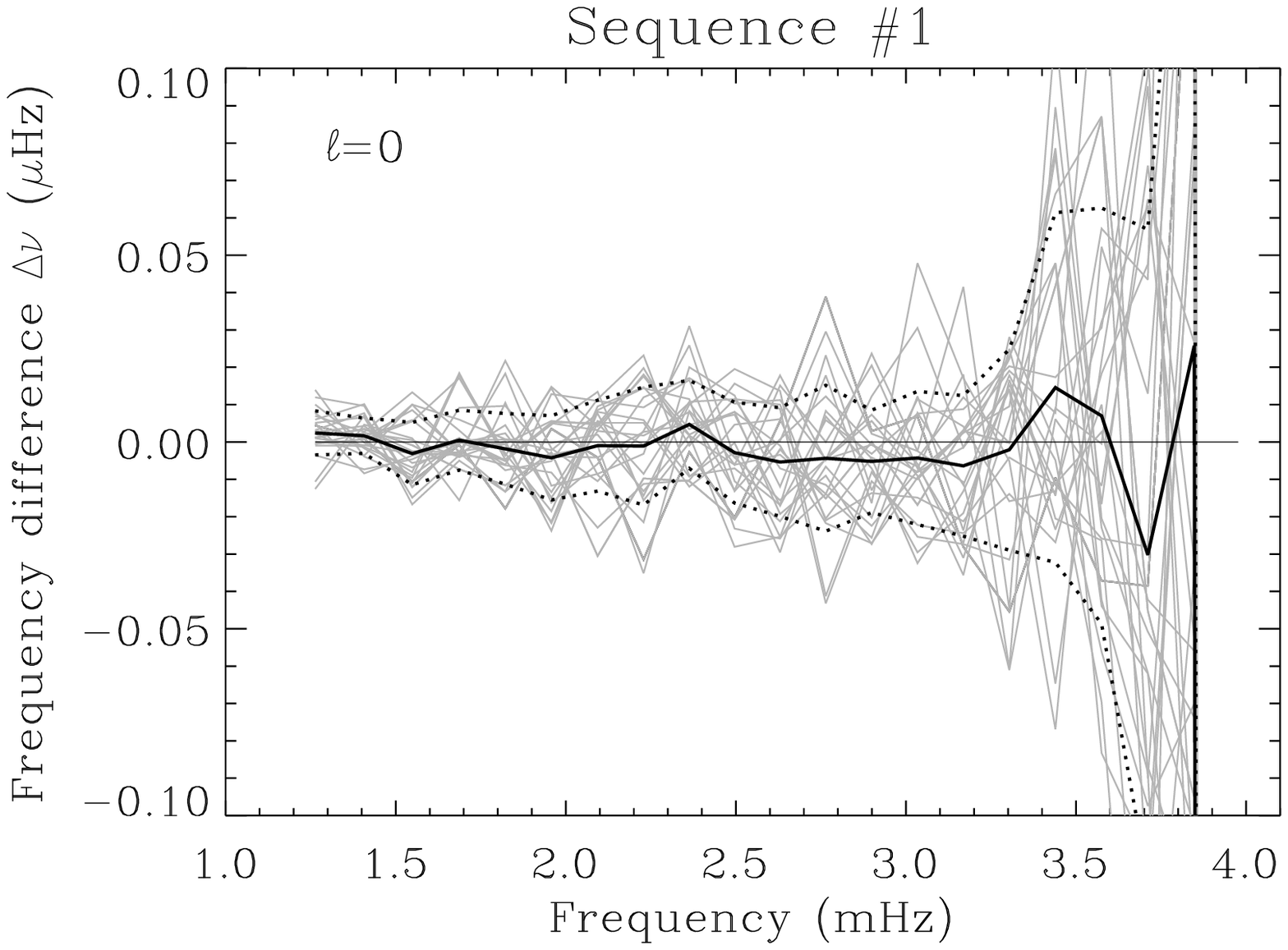}
               \epsfxsize=7.0cm\epsfbox{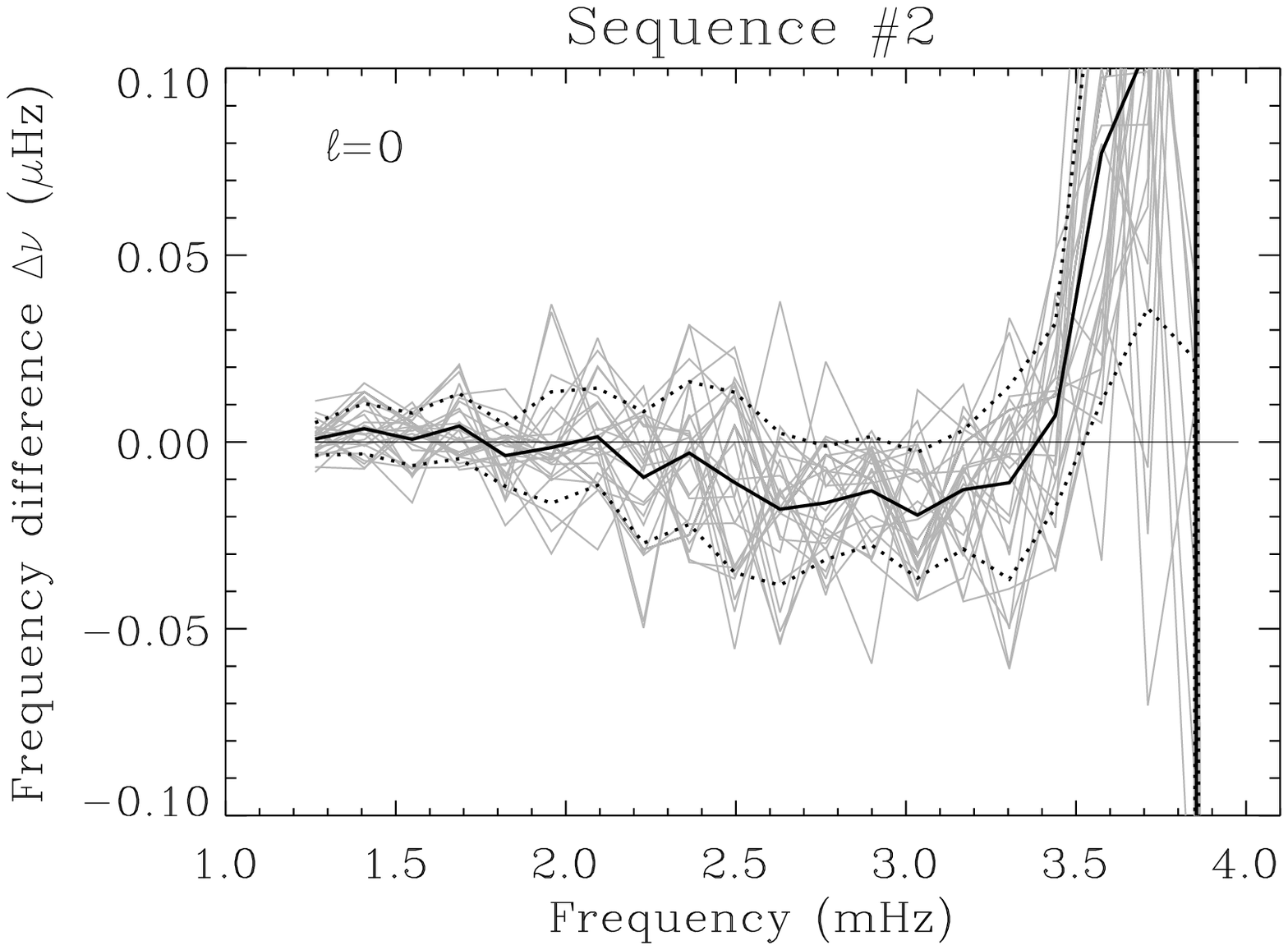}}
   \centerline{\epsfxsize=7.0cm\epsfbox{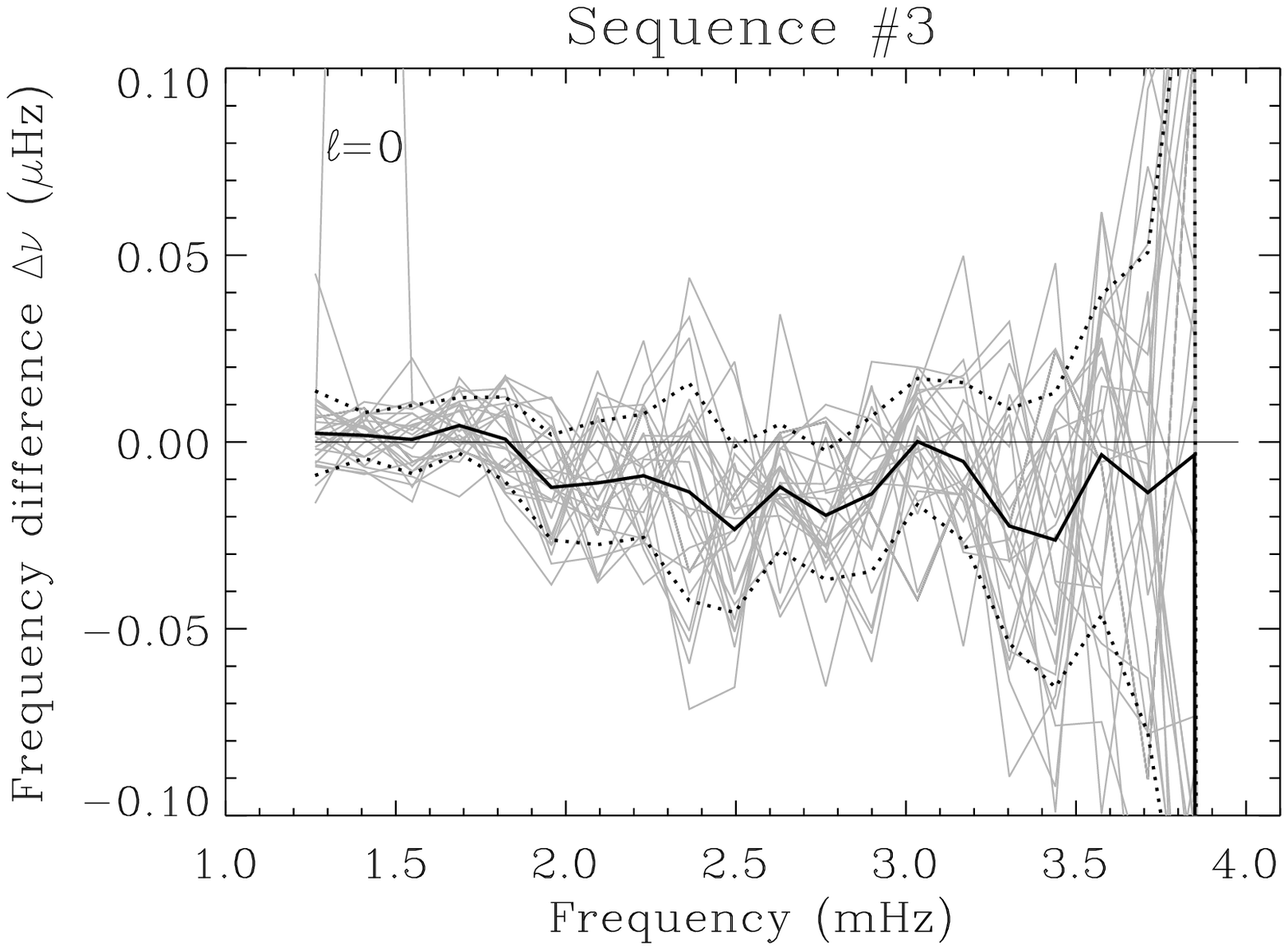}
               \epsfxsize=7.0cm\epsfbox{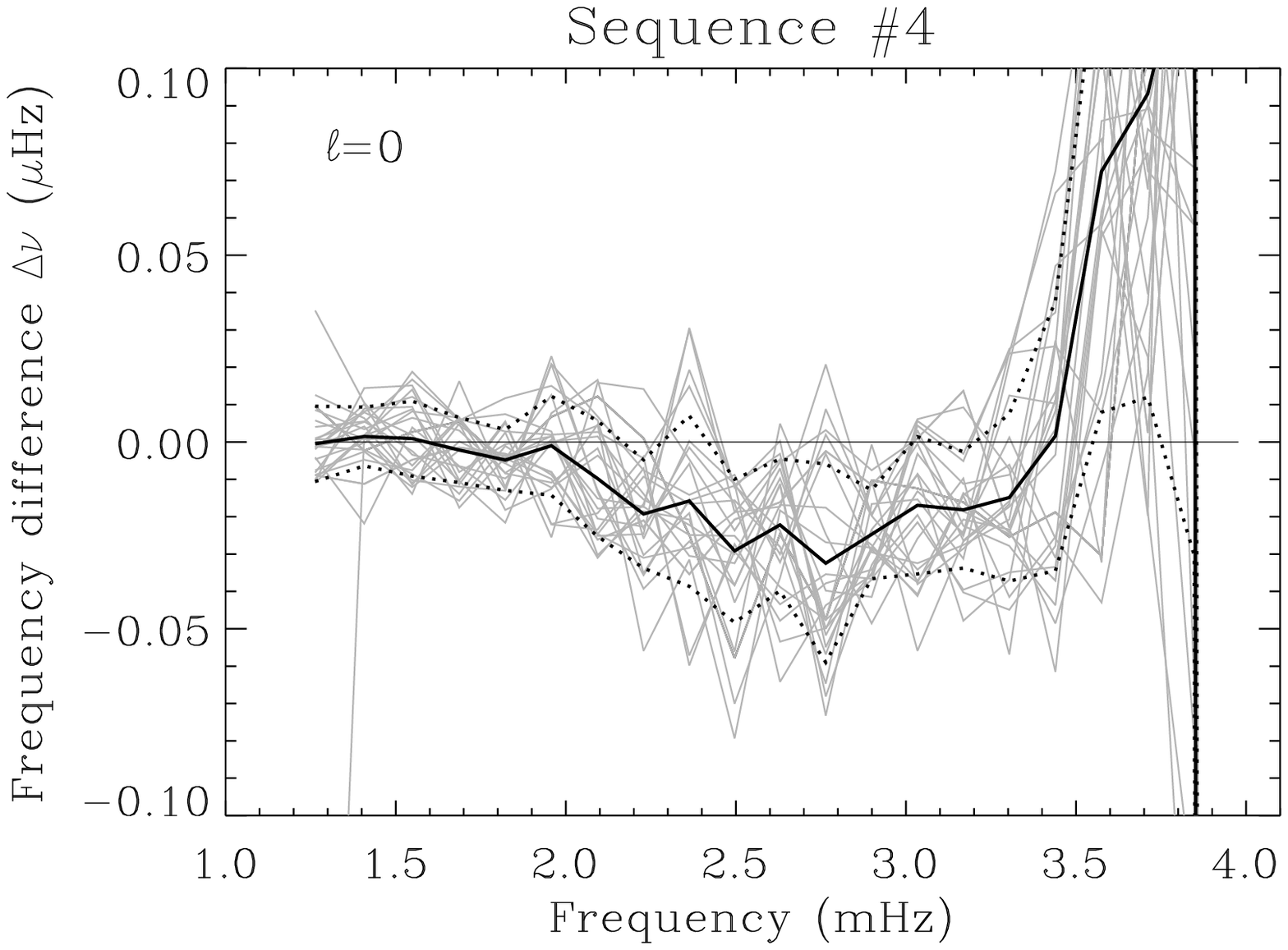}}

 \caption{Differences between the fitted and input $l=0$ frequencies,
 for fits to the four sequences of artificial solarFLAG datasets (see
 text for details).}

 \label{fig:nuds0}
 \end{figure*}

\begin{figure*}
   \centerline{\epsfxsize=7.0cm\epsfbox{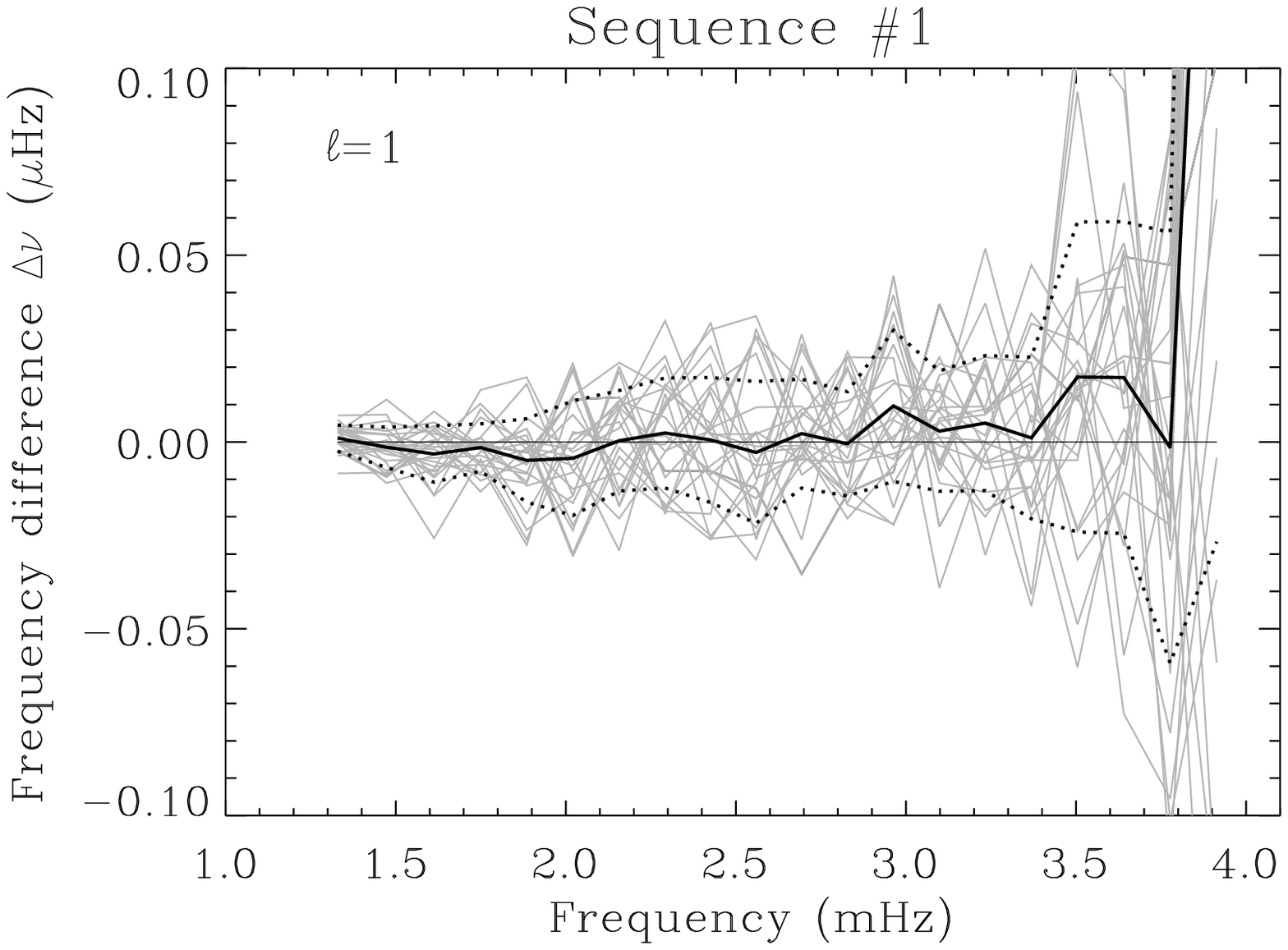}
               \epsfxsize=7.0cm\epsfbox{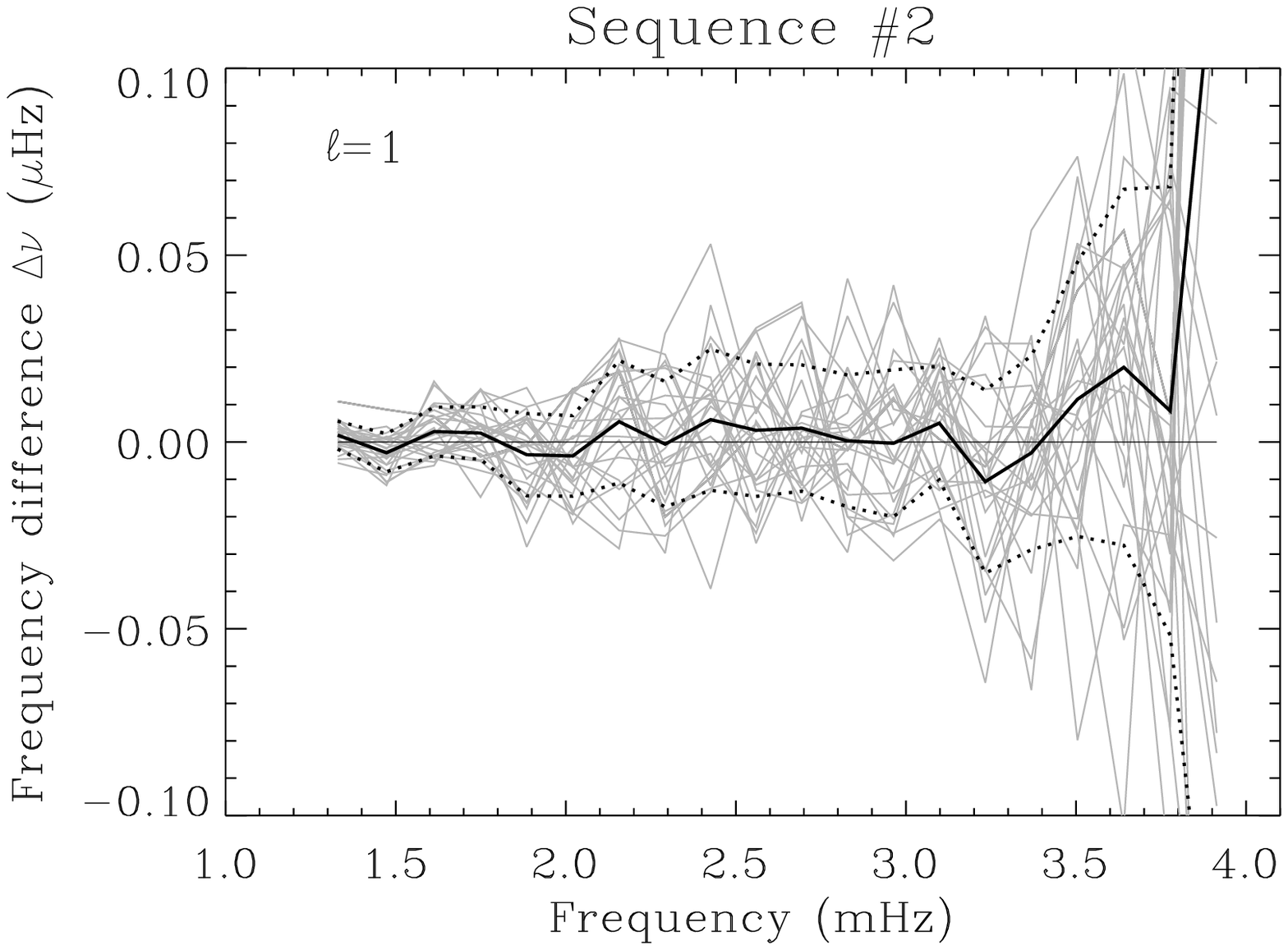}}
   \centerline{\epsfxsize=7.0cm\epsfbox{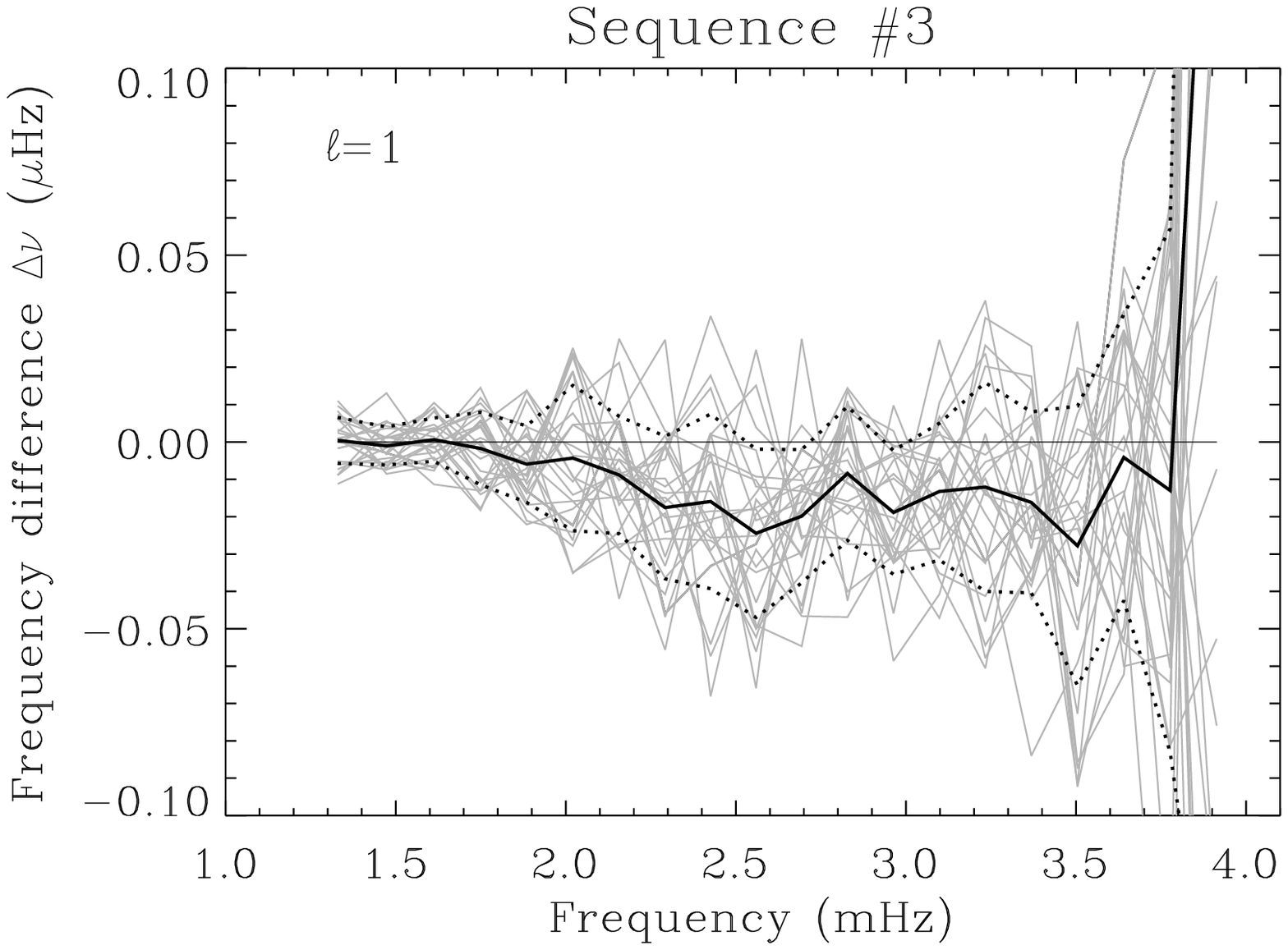}
               \epsfxsize=7.0cm\epsfbox{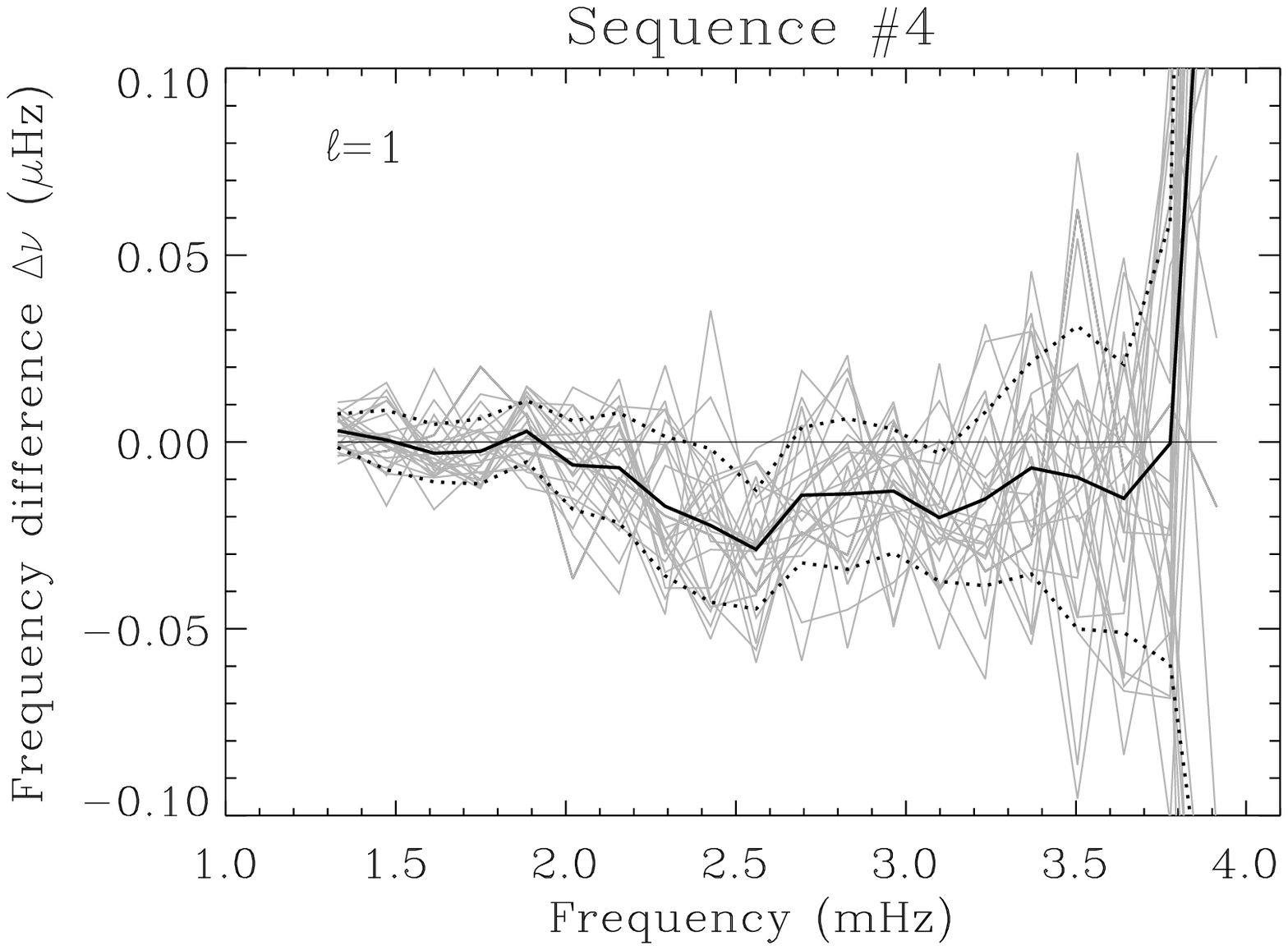}}

 \caption{Differences between the fitted and input $l=1$ frequencies,
 for fits to the four sequences of artificial solarFLAG datasets.}

 \label{fig:nuds1}
 \end{figure*}

Figs.~\ref{fig:nuds0} and~\ref{fig:nuds1} plot differences between the
fitted and input $l=0$ and $l=1$ frequencies, respectively, of all
four sequences. We selected these degrees to show the impact on the
even-$l$ and odd-$l$ pair fits, respectively.  Results on individual
datasets in each sequence are rendered in grey; the dark solid lines
show the average frequency differences for each sequence, while the
dotted lines bound the $1\sigma$ standard deviations on these average
differences.


\begin{figure*}
   \centerline{\epsfxsize=8.5cm\epsfbox{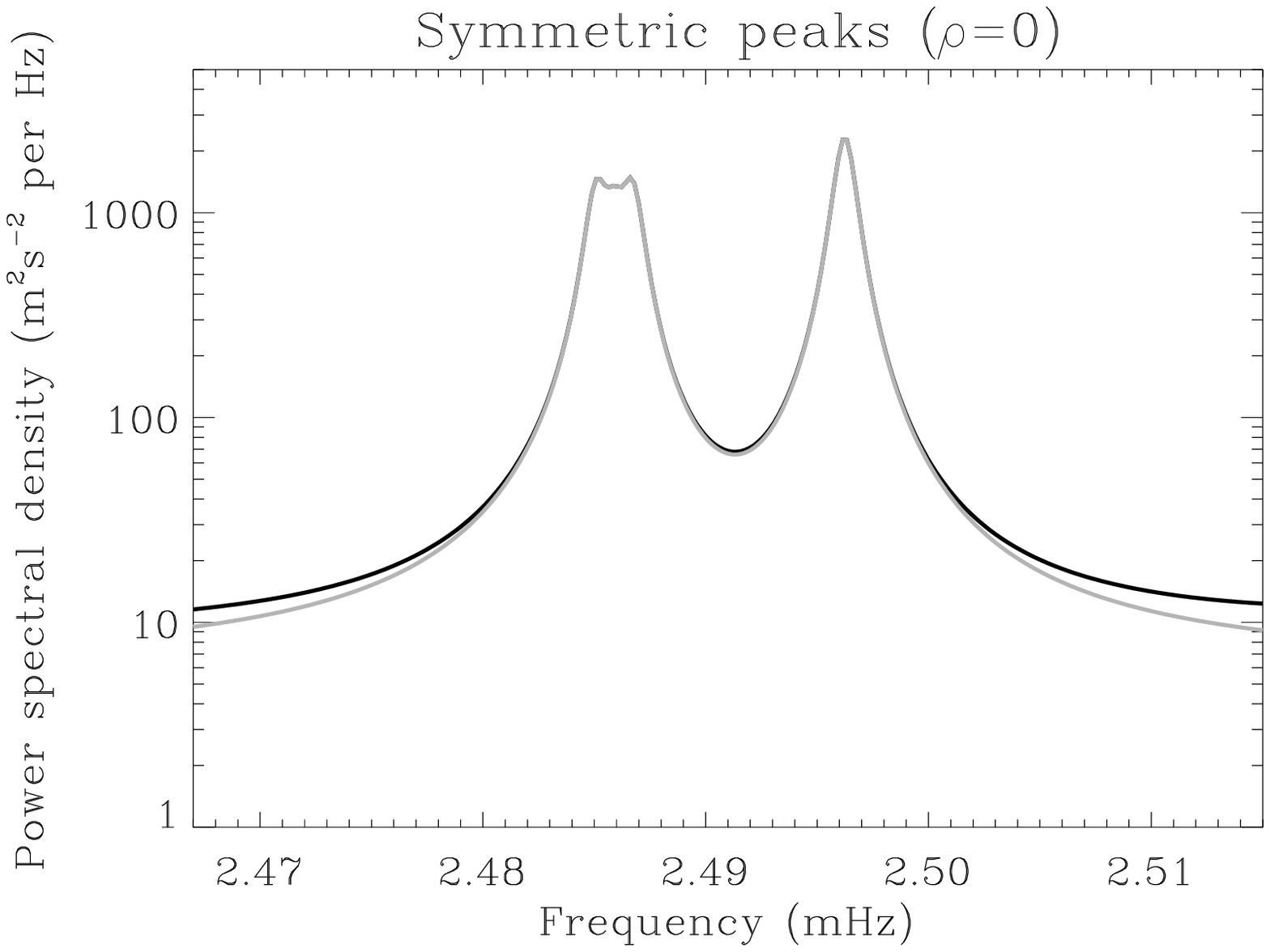}
               \epsfxsize=8.5cm\epsfbox{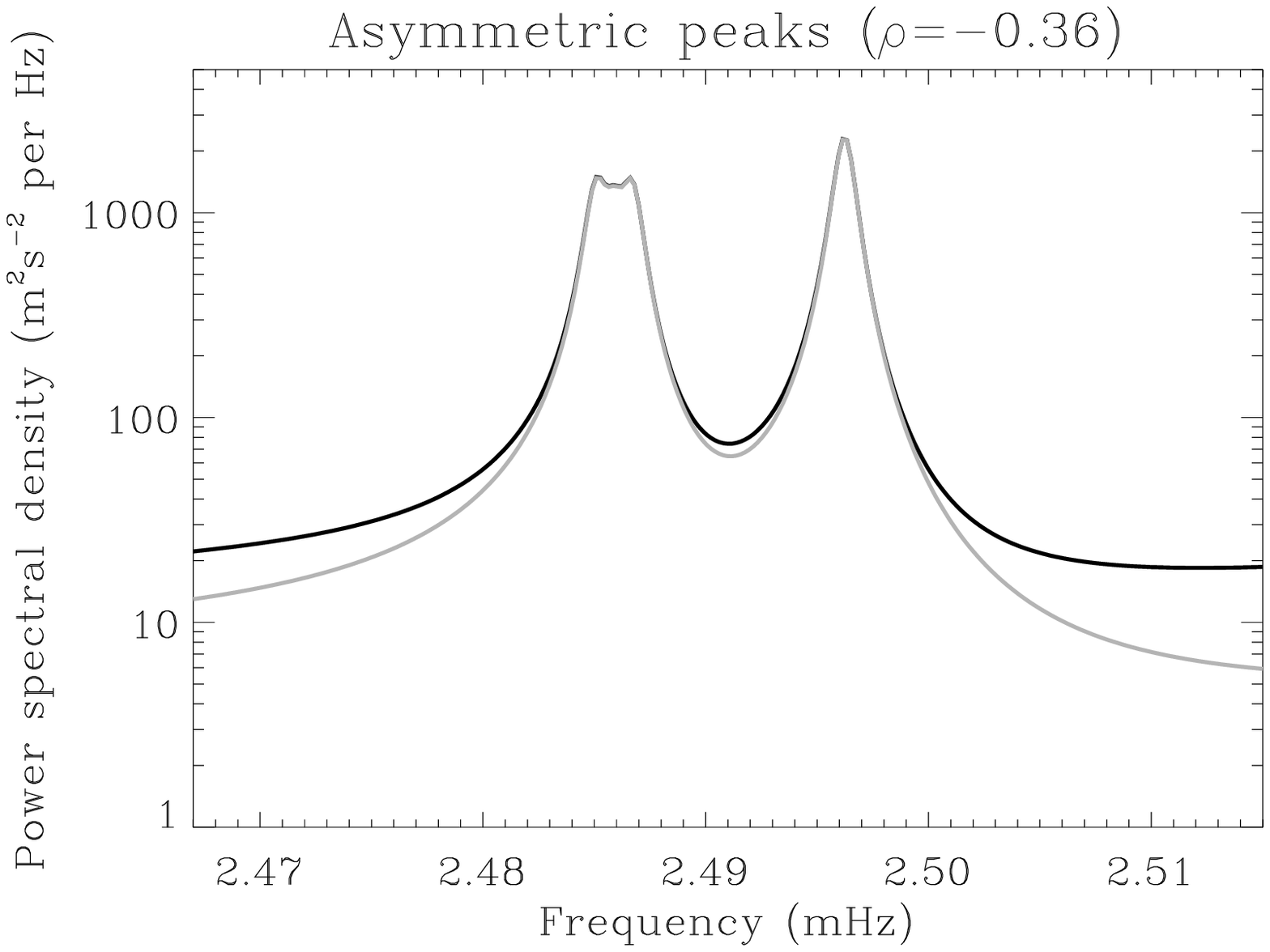}}

 \caption{The black curves show the limit frequency power spectra of
 Sequence\,\#1 (left-hand panel) and Sequence\,\#3 (right-hand panel),
 in the region of an $l=2$ ($2.486\,\rm mHz$), $l=0$ ($2.496\,\rm
 mHz$) mode pair. The grey curves show the power spectral density due
 only to the displayed $l=2$ and 0 mode pair, and the background
 noise.}

 \label{fig:pairs}
 \end{figure*}

The standard peak-bagging code evidently does a good job of recovering
the input frequencies when it is presented with the Sequence\,\#1 data
(upper left-hand panels of Figs.~\ref{fig:nuds0}
and~\ref{fig:nuds1}). There are no $l=4$ and 5 modes to give problems
for the fits; and because there are no correlations in the data, all
mode peaks are symmetric. Furthermore, even though the fitting did not
take account of outlying power from all other modes, the results were
not affected adversely. We shall come back to this point below
(discussion of Fig.~\ref{fig:pairs}), where we show that matters are
not so simple when the mode peaks are asymmetric.

We draw an important conclusion from the Sequence\,\#1 results:
provided the mode peaks are symmetric, failure to include power from
outlying modes [source (ii)] \emph{will not} bias the estimated
frequencies.

The upper right-hand panels of Figs.~\ref{fig:nuds0}
and~\ref{fig:nuds1} show the results for Sequence\,\#2. These datasets
now included the $l=4$ and 5 modes, but, again, peaks were
symmetric. The $l=0$ frequencies are seen to be biased, because the
standard peak-bagging failed to take account of power from the newly
introduced $l=4$ and 5 modes. The $l=1$ frequencies remained
unaffected, because the $l=4$ and 5 modes did not give a significant
contribution to the power spectral density in their fitting
windows. So, we may draw another important conclusion, this time from
the Sequence\,\#2 results: failure to account for power from the $l=4$
and 5 modes [source (i)] will bias estimates of even-$l$
frequencies. The size of this bias will depend on the visibilities of
the $l=4$ and 5 modes, relative to their more prominent $l=0$ and 2
counterparts, and the width in frequency of the fitting windows (wider
windows will admit more power from the $l=4$ and 5 modes). For the
typical standard peak-bagging scenario tested here -- even-$l$ fitting
windows were 48-$\rm \mu Hz$ wide -- bias was present in the range
$\approx 2.2$ to $\approx 3.4\,\rm mHz$. On average, the bias reached
sizes comparable to the estimated frequency uncertainties. However, in
some isolated cases (see individual fits shown as grey curves) the
bias could be up to three-times as large as the uncertainties.

Similar results were given for fitting window widths of between 40 and
$50\,\rm \mu Hz$ (the range covered by the ten hounds). If the windows
are made any narrower -- the simplest approach to reducing the impact
of the $l=4$ and 5 modes -- a new bias is introduced. This bias
appears because the windows are then too narrow to get robust
estimates of the falling power in the wings of the mode peaks. If the
windows are instead made wider, the impact of the $l=4$ and 5 modes on
the fitted frequencies becomes more severe. For example, when fitting
windows were widened to $70\,\rm \mu Hz$, the bias reversed sign above
$\approx 2.5\,\rm mHz$, and was found to be more than four-times as
large as the bias given with 40-$\rm \mu Hz$ windows.

Let us now turn to the Sequence\,\#3 and Sequence\,\#4 datasets, which
both included the effects of correlations and therefore had asymmetric
mode peaks. Results from the Sequence\,\#3 data (lower left-hand
panels of Figs.~\ref{fig:nuds0} and~\ref{fig:nuds1}) show biased $l=0$
and $l=1$ frequencies. The Sequence\,\#3 datasets contained no $l=4$
and 5 modes, so the source (i) bias could not have been a
factor. Rather, it is the source (ii) bias that now comes into
play. In summary: failure to account for the power due to modes
outside the fitting windows matters when the peaks were
asymmetric. Recall it did not matter when the peaks were symmetric
(see discussion on Sequence\,\#1 above). To help explain these
conclusions, consider Fig.~\ref{fig:pairs}.


The panels in this figure each show the limit frequency power spectrum
(black curves) for a different scenario. The left-hand panel shows the
limit spectrum for the case of Sequence\,\#1, where all peaks are
symmetric Lorentzians. The right-hand panel shows the limit spectrum
for the case of Sequence\,\#3, where the introduction of correlations
made the peaks asymmetric. The same narrow range in frequency has been
chosen for both plots. This range corresponds to the frequency fitting
window that would be selected to perform a standard (pair-by-pair) fit
to the $l=2$ (left-hand mode), $l=0$ (right-hand mode) pair shown.

We recall that in the standard fitting approach, the fitting model
includes only power from the mode pair in the chosen window (plus a
background term). The grey curve in each panel of Fig.~\ref{fig:pairs}
shows this power, i.e., the power spectral density due only to the
displayed modes and the background noise.  The obvious shortcoming of
the standard fitting approach is then made apparent by comparing the
black and grey curves in both panels: one is attempting to fit the full
spectrum (black curve) using a fitting model which is instead
correctly represented by the grey curve.

The difference between the black and grey curves in either panel of
course gives the contribution of power from the outlying modes [i.e.,
source (ii)]. We see that for Sequence\,\#1 the curves are
indistinguishable in the immediate neighbourhood of the
peaks. Furthermore, the mismatch of power further out looks very
similar at the low- and high-frequency ends of the window. We might
therefore expect standard fitting estimates of the frequencies from
Sequence\,\#1 to not be affected significantly by failing to model the
outlying power; and this is of course what we saw in the Sequence\,\#1
fitting results (see above).

Mismatches in power for the asymmetric Sequence\,\#3 data are, in
contrast, very evident in the vicinity of the peaks. Furthermore, the
mismatches have different sizes at the low- and high-frequency ends of
the window. This presents problems for the standard peak-bagging,
which tries to fit the black curve to a model comprising only the two
displayed modes (plus background), when it is of course the \emph{grey
curve} that actually describes the power of displayed modes.

There is more power in the grey curve at the low-frequency end of the
window compared to the high-frequency end, because the modes have
negative asymmetry. However, that difference in power is less
pronounced in the black curve. Attempts to fit the black curve to a
model comprising the two modes will therefore tend to return estimates
of the asymmetry that are smaller, and less negative, than the true
asymmetry, the latter seen in the shape of the grey curve. (Note that
the best-fitting estimate of the observed power will be close to the
black curve.)  The estimated asymmetries will therefore be positively
biased. This will in turn lead directly to negative bias in the
estimated frequencies.

These conclusions are borne out by checking the fitted asymmetries,
and fitted frequencies, of the Sequence\,\#3 results. The predicted
positive bias is present in both the $l=0$ and $l=1$ asymmetry
estimates (see also Fig.~\ref{fig:ass}). Furthermore, we see the
expected (highly significant) anti-correlation between bias in the
asymmetries and bias in the frequencies.

We may also use the plots in Fig.~\ref{fig:pairs} to explain why two
of the hounds who tried to allow for outlying power still obtained
frequencies that were biased. Both hounds modelled the outlying power
in terms of symmetric Lorentzians, while the hare-and-hounds data of
course contained asymmetric mode peaks. The left-hand panel of
Fig.~\ref{fig:pairs} shows that modelling the outlying power in this
way will lead to only fairly modest changes in the power spectral
density. We would therefore have expected the ``outlying'' power in
the two hounds' fitting models to have made little difference in fits
to asymmetric hare-and-hounds data. Looking at the right-hand panel of
Fig.~\ref{fig:pairs}, their model representations of the two modes
plus outlying power would have been little different to the grey curve;
and so the estimated frequencies remained biased.

We draw perhaps the most important conclusion of the paper from the
Sequence\,\#3 results: failure to account for power from modes whose
central frequencies lie outside the fitting windows will bias
estimates of both the even-$l$ and odd-$l$ frequencies, above $\approx
1.8\,\rm mHz$, by an amount that, on average, can again reach the size
of the typical frequency uncertainties. In some isolated cases (see
individual fits shown as grey curves) the frequency bias may be up to
three-times as large as the uncertainties.

Finally, the Sequence\,\#4 results (lower right-hand panels of
Figs.~\ref{fig:nuds0} and~\ref{fig:nuds1}) bring everything together,
and show the total impact of our two main sources of bias. Recall
these datasets contained $l=4$ and 5 modes, and correlations, and as
such they had the same underlying properties as the hare-and-hounds
dataset. The results demonstrate that the total frequency bias is on
average most significant in the estimated $l=0$ frequencies, because
source (i) and source (ii) give a similar-sized contribution to the
bias. In some isolated cases (again, see the grey curves) the total
bias may be almost four-times the size of the frequency
uncertainties. The bias is on average less severe in the $l=1$
results, because only the source (ii) bias plays a significant r\^ole
in affecting the odd-$l$ pair fits.

 \section{Summary and Concluding Discussion}
 \label{sec:sum}

We have used the new solarFLAG simulator, which includes the effects
of correlated mode excitation and correlations with background noise,
to make artificial timeseries data that mimic Doppler velocity
observations of the Sun as a star. The correlations give rise to
asymmetry of mode peaks in the frequency power spectrum.

A 3456-day dataset was used as the input data for the latest solarFLAG
hare-and-hounds exercise. This paper reports on the results of that
exercise, which was concerned with testing methods for extraction of
p-mode frequencies of low-degree (low-$l$) modes. Ten hounds applied
their peak-bagging codes to the hare-and-hounds dataset. Peak-bagging
involves maximum-likelihood fitting of mode peaks in the frequency
power spectrum to multi-parameter fitting models.  Each hound returned
peak-bagging estimates of the frequencies of the artificial $l=0$ to 3
modes to the hare (who was WJC) for further scrutiny.

Analysis of the results showed clear evidence of a systematic bias in
the estimated frequencies of modes above $\approx 1.8\,\rm mHz$. The bias
is negative, meaning the estimated frequencies systematically
underestimate the input frequencies. A follow-up analysis on
independent realizations of the hare-and-hounds dataset showed that in
some fits the bias could be as much as three- to four-times as large
as the frequency uncertainties. Over the affected range of mode
frequencies, the average bias is typically one to two-times the
frequency uncertainties.

We identified two sources that are the dominant contributions to the
frequency bias. Both sources involve failure to model accurately
subtle aspects of the observed power spectral density in the part
(window) of the frequency power spectrum that is being fitted.  One
source of bias arises from a failure to account for the power spectral
density of the weak $l=4$ and 5 modes. The other source arises from a
failure to account for the power spectral density coming from all
those modes whose frequencies lie outside the fitting windows
(``outlying'' power).

The main lesson to be drawn from this paper is that the Sun-as-a-star
peak-bagging codes need to allow for both sources, otherwise the
frequencies given by analysis of real Sun-as-a-star data will in all
likelihood be biased. The identification, and measurement, of the bias
from ``outlying'' power is the most important new finding of the
paper. Can we afford to ignore its effects? The short answer must be
no. Our analysis suggests the magnitude of its frequency bias may be
up to three-times as large as the frequency uncertainties, depending
on the impact of realization noise, and it could present problems for
helioseismic inference on the solar structure from inversions of the
mode frequencies.

The precise sizes of biases given on the real Sun-as-a-star data will
clearly depend on how closely our artificial data resemble those real
data. We are certainly now able to reproduce frequency power spectra
that bear a close resemblance to the real spectra -- courtesy of the
new solarFLAG simulator -- and this suggests our bias estimates do
have quantitative merit where helioseismic predictions concerning the
real data are concerned. However, we should bear in mind that there
may be some aspects of the real Sun-as-a-star data that are not
reproduced exactly in the artificial data.

One such detail concerns the exact shapes shown by the asymmetric mode
peaks. In the real solar p-mode data, it is assumed that there is also a
contribution to the asymmetry owing to the radial location, extent,
and multipole properties, of the acoustic sources. The question then
arises: Is the underlying form of the power spectral density due to
these contributions the same as that from the correlated noise
modelled in the solarFLAG simulator? Subtle differences would affect
the sizes of the frequency biases.

We finish with a few comments on implications of the results of this
paper for the peak-bagging codes.  The standard Sun-as-a-star approach
is to go through the frequency power spectrum fitting a pair of modes
at a time. For such an approach to be useful our results here stress
the need for power from the outlying modes to be fully accounted for
in the fitting windows (so-called ``pseudo whole-spectrum''
fitting). The other option is to fit all modes in the frequency power
spectrum in one go (so-called ``whole spectrum'' fitting). Either way,
it is not sufficient to have an approximate, or first-order, estimate
of the outlying, or total, power spectral density: the estimate must
be very accurate, otherwise the frequency bias will remain, or
additional bias may be introduced.

In order to provide such an estimate, it is necessary to describe
accurately the power spectral density of each mode peak a long way
from its resonant frequency (i.e., in the decaying wings of the
peaks). The fitting formalism most often used to model the asymmetric
power spectral density of the mode peaks -- that due to Nigam \&
Kosovichev (1998) -- fails in such a description. This is because the
Nigam \& Kosovichev formalism is an approximation (low-order
expansion) that is usable only at frequencies close to resonance. Far
from resonance, the modelled power spectral density tends to a
constant offset not shown by the real data\footnote{The ``Fano
profile'' formula in Gabriel et al. (2001) also fails in this regard,
because it retains those terms that lead to the offset far from
resonance.}

Clearly the requirements on the sought-for fitting model are that it
should describe the asymmetric peaks both close to resonance, and far
from resonance where we know the power falls off significantly in the
real p-mode peaks.  There is, potentially, a very simple solution to
this problem. If the high-order terms in the Nigam \& Kosovichev
formalism (those in the square of the asymmetry parameter) are
disregarded, it turns out that the resulting, truncated formalism can
satisfy both of the above requirements. Indeed, it has the correct
form to model accurately the asymmetric shapes given by correlated
noise (see Toutain, Elsworth \& Chaplin 2006). The issue then arises
as to whether this is sufficient to describe the shapes of the real
p-mode peaks (see previous comments above). Another approach is to
generate a model of resonant spectrum in one go, with all the overtone
structure included, rather than model the spectrum as a superposition
of individual Lorentzian-like peaks. This may be accomplished using a
suitable form for the acoustic potential of the solar cavity, together
with information on the acoustic source and the reflection and
transmission properties of the upper cavity boundary (e.g., see
Jefferies, Vorontsov \& Giebink 2004).

Finally, we note that the detailed conclusions drawn in this paper are
for the Sun-as-a-star observations. The Sun-as-a-star techniques are
of course also directly applicable to stars that show Sun-like
oscillations, and we are now starting to get \emph{asteroseismic}
peak-bagging results on other stars (e.g., see Fletcher et al. (2006),
who analyzed data collected by WIRE on $\alpha$\,Cen~A; and
Appourchaux et al. (2008), who have analyzed the first Sun-like
oscillations data collected by CoRoT, on the star HD49933). A
peak-bagging pipeline is being constructed by the asteroFLAG
group\footnote{http://www.issibern.ch/teams/Astflag} (Chaplin et
al. 2008a) for application to the asteroseismic data that will be
collected on hundreds of Sun-like stars by the NASA Kepler mission
(Christensen-Dalsgaard 2007). Even though the intrinsic
signal-to-noise ratios will be lower than for the Sun-as-a-star data
some stars will be monitored continually for several years, meaning we
should be able to constrain the p-mode parameters to high levels of
precision.  With the p-mode parameters reflecting different intrinsic
properties of the stars, we should expect to be confronted with many
different potential bias problems in different parts of the
color-magnitude diagram occupied by the Sun-like oscillators
(Appourchaux et al. 2006a, b; Chaplin et al. 2008b). Similar studies
to the one undertaken here for the Sun as a star are now being made by
asteroFLAG to prepare the peak-bagging analysis for Kepler.

\subsection*{ACKNOWLEDGMENTS}

WJC and SJJ-R acknowledge the support of STFC. This work was also
supported by the European Helio- and Asteroseismology Network (HELAS),
a major international collaboration funded by the European
Commission's Sixth Framework Programme. WJC also thanks Sarbani~Basu
and Aldo~Serenelli for providing the model BS05(OP) frequencies.

\end{document}